\documentclass[eqsecnum,aps,prd,nofootinbib]{revtex4}
\usepackage{graphicx}

\def\ben{\begin{equation}}
\def\een{\end{equation}}
\def\bena{\begin{eqnarray}}
\def\eena{\end{eqnarray}}

\usepackage{amsmath}
\usepackage{enumerate}

\begin{document}

\title {On the fate of black string instabilities: An Observation}

\author{Donald Marolf}

\affiliation{Physics Department, UCSB, Santa Barbara, CA 93106, USA\\
{\tt marolf@physics.syr.edu}}

\date{April, 2005}

\begin{abstract}
Gregory and Laflamme (hep-th/9301052) have argued that an instability causes the Schwarzschild black string to break up into disjoint black holes.  On the other hand, Horowitz and Maeda (arXiv:hep-th/0105111) derived bounds on the rate at which the smallest sphere can pinch off, showing that, if it happens at all, such a pinch-off can occur only at infinite affine parameter along the horizon.  An interesting point is that, if a singularity forms, such an infinite affine parameter may correspond to a finite advanced time -- which is in fact a more appropriate notion of time at infinity.  We argue below that pinch-off at a finite advanced time is in fact a {\it natural} expectation under the bounds derived by Horowitz and Maeda.
\end{abstract}
\maketitle

\section{Introduction}

Although black holes in 3+1 dimensions are largely stable, an
intriguing instability of Schwarzschild black strings in 4+1
dimensions was discovered by Gregory and Laflamme \cite{GL,GL2}
and has been the focus of numerous investigations
\cite{HM,SG,Morse,TW,TW2,ES,KS,Park,TW3,KRev,HORev,C,GLP}, both
analytic and numerical.  Perhaps the most interesting issue is the
question of the end-point of time evolution under this
instability. Gregory and Laflamme originally argued on entropic
grounds that the black string should break up into a set of
disconnected black holes. Such behavior is particularly
interesting as it would provide a counter-example to  at least
certain versions of the Cosmic Censorship Conjecture in 4+1
dimensions.

The details of the pinch-off process were considered by Horowitz and Maeda \cite{HM}, who used the Raychaudhuri equation \cite{Wald} to derive bounds on how fast the area-radius $R$ of smallest sphere on the horizon can shrink in terms of the affine parameter $\lambda$ along null generators of the horizon.  They found that if $R$ tends to zero, it must do so quite slowly with $R$ vanishing only at infinite affine parameter.  Imposing the physically reasonable assumption that the final horizon area remains finite\footnote{Actually, \cite{HM} derives the stated restrictions from the stronger requirement that the integrated expansion $\int^\infty \theta(\lambda) d \lambda$ converge on the `neck'  where the black string pinches off.  This requirement is also physically reasonable, but is technically stronger than finiteness of the total horizon area (which would also involve an integral along the string direction of the horizon).} further restricts the analytic form of $R(\lambda):$   power laws and simple exponential behaviors are forbidden, and for functions of the form
\begin{equation}
\label{HMbound}
R = A e^{-(\ln \lambda)^{\alpha}}
\end{equation}
with constant $A$, one must have $0 < \alpha < 1/2$.

However, as mentioned in \cite{GLP}, delaying the pinch-off until infinite affine parameter along the horizon is not necessarily the same as delaying the pinch-off until an infinite value of some time measured near infinity.  Thus, the Horowitz-Maeda  argument does not rule out a scenario in which the black string breaks up into black holes at a finite time as measured by an asymptotic observer\footnote{The author first heard this suggestion from Robert M. Wald, in a conversation at the GR17 conference in Dublin, Ireland, July 2004.}.  If a singularity forms on the horizon, then the relationship between the affine parameter on the horizon and some asymptotic time is allowed to become singular.

Our purpose here is to note that such a scenario is a natural expectation based on the bounds derived by Horowtiz and Maeda.  In particular, we show below that if {\it i}) the decay of the radius follows  (\ref{HMbound}) for some $A$, $0< \alpha < 1/2$ and {\it ii}) the relationship between affine parameter and advanced time at infinity can be modeled on the familiar relationship for static black holes, then the advanced time remains finite at $\lambda = \infty.$  Such a scenario would resolve an apparent tension between the results of \cite{HM} and various investigations \cite{SG,Morse,TW,TW2,ES,KS,TW3,KRev} which together indicate that, at least in spacetime dimension $d \le 13$, the only equilibrium final state to which the instability could lead is a set of separated black holes.  This argument is given in section \ref{main} below, after which we provide a brief discussion (section \ref{disc}).

\section{Affine parameter vs. Advanced Time}
\label{main}

Both \cite{GL} and \cite{HM} considered the fate of the Schwarzschild black string in 4+1 dimensional Einstein-Hilbert gravity. This black string spacetime is the direct (metric) product of a spacelike real line with the 3+1 Schwarzschild solution.  In familiar coordinates, the exterior is described by the line element:
\begin{equation}
\label{SBS}
ds^2 = -(1-R/r) dt^2 + \frac{dr^2}{1-R/r} + r^2d\Omega^2 + dz^2,
\end{equation}
where $d\Omega^2 = d\theta^2 + \sin^2 \theta d\phi^2$ is the line element on the unit two-sphere and $R$ is a constant giving the value of the area-radius $r$ at the horizon.

\begin{figure}
  \centering
    \includegraphics[width=6cm]{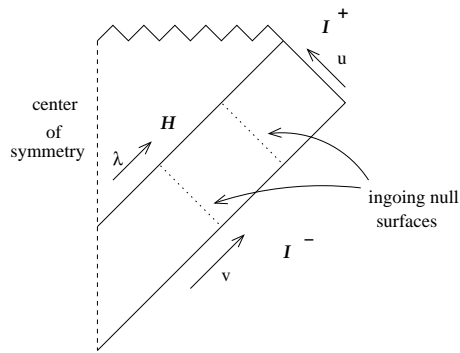}
  \caption{A conformal diagram is shown for the case where the black string forms from gravitational collapse (so that there is a regular center of the rotational symmetry).  Past and future null infinity $({\cal I}^\pm)$ are shown, along with the horizon $H$.  The affine parameter $\lambda$ along $H$ and the advanced time $v$ along ${\cal I}^-$ are indicated, and can be compared on ingoing null surfaces (dotted lines).   A similar retarded time $u$ can of course also be introduced along ${\cal I}^+$.}
 \end{figure}

Many salient features of this spacetime may be described by a two-dimensional conformal diagram similar to that associated with the Schwarzschild black hole.  Figure 1 shows this diagram for a related spacetime where the black string forms from gravitaitonal collapse (so that there is a regular center of the rotational symmetry).  The diagram  indicates the affine parameter $\lambda$ along the horizon, as well as the advanced time $v$ along past null infinity.  Note that, while both $\lambda$ and $v$ are defined as parameters along distinct null curves ($\lambda$ on the horizon and $v$ at infinity), either may naturally be taken to label the family of ingoing null surfaces (dotted lines).

Similar parameters may be introduced in any static asymptotically flat spacetime where the time-translation has a Killing horizon.  If the horizon is non-degenerate one has \cite{Wald}
\begin{equation}
\label{rel}
\frac{d \lambda}{\lambda} = \kappa \ dv,
\end{equation}
where $\kappa \neq 0$ is the surface gravity of the timelike Killing vector field.  In the particular case of Schwarzschild (\ref{SBS}), we have $\kappa = \frac{1}{2R}.$

Here we wish to consider the relationship between $\lambda$ and $v$ for the dynamic spacetime that results from activating the Gregory-Laflamme instability \cite{GL} of the Schwarschild black string. In particular, we follow \cite{HM}  in considering a perturbation which has both $SO(3)$ rotation invariance around the black string and reflection symmetry about some hypersruface $z=0$.  Thus, $z=0$ is a totally geodesic sub-manifold, and it is useful to draw a conformal diagram for this hypersurface even in our dynamic spacetime.  Furthermore, we can foliate the $z=0$ plane with null surfaces generated by radial null geodesics, much like the dotted lines shown in figure 1 for the static black string.  Thus, in the $z=0$ plane it is again meaningful to compare the affine parameter $\lambda$ along the horizon with the advanced time $v$ at infinity.  We may also speak of the radius $R$ of the horizon at each $\lambda$ (equivalently, at each $v$).

The relationship between $\lambda$ and $v$ may be quite complicated as there is no longer a time-translation symmetry.  Even on a null surface labeled by fixed $\lambda$ and $v$, the spacetime will not match any static black string due to the presence of both non-vanishing derivatives and radiation.  However, we are free to define an {\it effective} surface gravity $\kappa_{eff}$ such that a relation of the form (\ref{rel}) continues to hold:
\begin{equation}
\label{rel2}
\frac{d \lambda}{\lambda} = \kappa_{eff} dv.
\end{equation}
This effective surface gravity will of course be a function of $\lambda$ (or, equivalently, a function of $v$).  As further motivation for this definition, let us recall that \cite{GLP} analyzed the dynamical black string simulation of \cite{C} and found that evolution even of quantities associated with the horizon was more naturally described in terms of $\ln \lambda$ than in terms of $\lambda$.

While we have no analytic control over $\kappa_{eff}$, it is interesting to investigate the consequences of a plausible assumption:  namely, that $\kappa_{eff}$ scales with $R$ in a manner similar to the relationship for static black holes.  That is, we suppose that for Schwarzschild black branes (in any dimension) we have $\kappa_{eff} \ge k/R$ for some $k$.  Since Horowitz and Maeda noted that relations of the form (\ref{HMbound}) with $0 < \alpha < 1/2$ are consistent with their bounds, let us also suppose that such a relation holds.  Then, introducing $\beta = \ln \lambda$ we have
\begin{equation}
dv = \frac{d \ln \lambda}{\kappa_{eff}} \le \frac{R}{k} d \beta = \frac{A}{k} e^{-\beta^\alpha} d \beta.
\end{equation}
Integrating this relation and using $\alpha > 0$ we see that the advanced time $v$ asymptotes to a finite value as $R \rightarrow 0$ (i.e., as $\beta \rightarrow \infty$).

\section{Discussion}
\label{disc}

We have considered a plausible assumption for the scaling of the effective surface gravity (\ref{rel2}) at the neck (i.e., the smallest sphere) of the horizon of a dynamic black string.  In particular, by assuming that it scales with the neck size $R$ in the same manner as the surface gravity of a static black string scales with the horizon size,  we have shown that the bounds of Horowitz and Maeda \cite{HM} naturally lead to a scenario in which the neck reaches zero size at finite advanced time.  In this scenario, such an event occurs only at  {\it infinite} affine parameter along the horizon.  However, we note that it is the advanced time which is more directly relevant to an asymptotic observer.  In fact, it is plausible that any observer maintaining a finite separation from the horizon crosses the $R=0$ ingoing null surface at a finite proper time.  As noted in \cite{GLP}, such a scenario would resolve the apparent tension between \cite{HM} and detailed analytic and numerical investigations \cite{SG,Morse,TW,TW2,ES,KS,TW3,KRev}.

While we have no direct argument for the above assumption, we note that similar results continue to hold under much weaker conditions.  Static Schwarzschild black branes satisfy
$\kappa \propto 1/R.$  But to obtain the above conclusion it is sufficient that the effective surface gravity satisfy $\kappa_{eff} \ge k /R^{ \gamma}$ for any constant $\gamma > 0$.  Thus, there is still room for the dynamic black string to differ significantly from the static black string without affecting the conclusion.

In practice, it will likely be left to numerical simulations to determine the asymptotic behavior of $\kappa_{eff}$.   An an initial step in this direction, let us consider the analysis of \cite{GLP}, using the data set of \cite{C}.  Their figure 1 shows what in our notation is $\beta=\ln \lambda$ vs. the advanced time $v$, and from the plot it is clear that they find $\kappa_{eff}$ to increase over the range of the simulation.  While the data set of \cite{C} does not appear to have sufficient resolution to test our conjecture in detail, this may prove to be a fruitful direction for future investigation.

Even under the above scenario, interesting questions remain with regard to the dynamics.  In particular, while a bifurcation of the horizon requires a violation of the Kaluza-Klein generalization of strong future asymptotic predictability \cite{HE}, a weaker notion of cosmic censorship (such as a Kaluza-Klein generalization of just future asymptotic predictabiliity) might perhaps continue to hold.  The essential difference between the two asymptotic predictability criteria is that the strong version requires the Cauchy surfaces to extend inside the horizon, whereas the other version only requires that the past of ${\cal I}^+$ be globally hyperbolic.   In addition we may note that,  even if the bifurcation occurs at finite advanced time, this does not rule out the possibility that any singularity might occur only at infinite {\it retarded time} ($u$) measured along ${\cal I}^+$.  If this is the case, then ${\cal I}^+$ would remain complete.

\begin{acknowledgments}
The author would like to thank David Garfinkle, Stefan Hollands, Mukund Rangamani, Bob Wald, and Toby Wiseman for the conversation that inspired this work.  He would also like to thank Gary Horowitz for numerous conversations on the fate of black string instabilities and the organizers of the GR17 conference in Dublin, Ireland for the stimulating atmosphere of the conference.
This work was supported in part by NSF grant PHY0354978 and
by funds from the University of California.
\end{acknowledgments}


\begin{thebibliography}{11}
\bibitem{GL}
R.~Gregory and R.~Laflamme, ``Black strings and p-branes are
unstable,'' Phys.\ Rev.\ Lett.\  {\bf 70}, 2837 (1993)
[arXiv:hep-th/9301052].


\bibitem{GL2}
R.~Gregory and R.~Laflamme, ``The Instability of charged black
strings and p-branes,'' Nucl.\ Phys.\ B {\bf 428}, 399 (1994)
[arXiv:hep-th/9404071].


\bibitem{HM}
G.~T.~Horowitz and K.~Maeda,
  ``Fate of the black string instability,''
  Phys.\ Rev.\ Lett.\  {\bf 87}, 131301 (2001)
  [arXiv:hep-th/0105111].


\bibitem{SG}  S.~S.~Gubser,
  ``On non-uniform black branes,''
  Class.\ Quant.\ Grav.\  {\bf 19}, 4825 (2002)
  [arXiv:hep-th/0110193].

\bibitem{Morse}
 B.~Kol,
  ``Topology change in general relativity and the black-hole black-string
  transition,''
  arXiv:hep-th/0206220.



\bibitem{TW} T.~Wiseman,
  ``Static axisymmetric vacuum solutions and non-uniform black strings,''
  Class.\ Quant.\ Grav.\  {\bf 20}, 1137 (2003)
  [arXiv:hep-th/0209051].

\bibitem{TW2}
T.~Wiseman,
  ``From black strings to black holes,''
  Class.\ Quant.\ Grav.\  {\bf 20}, 1177 (2003)
  [arXiv:hep-th/0211028].



\bibitem{ES}  E.~Sorkin,
  ``A critical dimension in the black-string phase transition,''
  Phys.\ Rev.\ Lett.\  {\bf 93}, 031601 (2004)
  [arXiv:hep-th/0402216].

\bibitem{Park} M.~I.~Park,
  ``The final state of black strings and p-branes, and the Gregory-Laflamme
  instability,''
  arXiv:hep-th/0405045.


\bibitem{KS}
B.~Kol and E.~Sorkin,
  ``On black-brane instability in an arbitrary dimension,''
  Class.\ Quant.\ Grav.\  {\bf 21}, 4793 (2004)
  [arXiv:gr-qc/0407058].

\bibitem{TW3}
H.~Kudoh and T.~Wiseman,
  ``Connecting black holes and black strings,''
  arXiv:hep-th/0409111.


\bibitem{KRev} B.~Kol,
  ``The phase transition between caged black holes and black strings: A
  review,''
  arXiv:hep-th/0411240.

\bibitem{HORev}  T.~Harmark and N.~A.~Obers,
  ``Phases of Kaluza-Klein black holes: A brief review,''
  arXiv:hep-th/0503020.



\bibitem{C} M.~W.~Choptuik, L.~Lehner, I.~Olabarrieta, R.~Petryk, F.~Pretorius and H.~Villegas,
 ``Towards the final fate of an unstable black string,''
  Phys.\ Rev.\ D {\bf 68}, 044001 (2003)
  [arXiv:gr-qc/0304085].



\bibitem{GLP} D.~Garfinkle, L.~Lehner and F.~Pretorius,
  ``A numerical examination of an evolving black string horizon,''
  Phys.\ Rev.\ D {\bf 71}, 064009 (2005)
  [arXiv:gr-qc/0412014].


\bibitem{Wald} R. M. Wald, {\it General Relativity} (University of Chicago Press, Chicago, 1984).

\bibitem{HE}  S. W. Hawking and G. F. R. Ellis, {\it The large scale structure of space-time} (Cambridge University Press, Cambridge, 1973).  See proposition 9.2.5.

\end{thebibliography}
\end{document}